# The Security Risk of Lacking Compiler Protection in WebAssembly

Quentin Stiévenart
Vrije Universiteit Brussel, Belgium
quentin.stievenart@vub.be

Coen De Roover
Vrije Universiteit Brussel, Belgium
coen.de.roover@vub.be

Mohammad Ghafari
University of Auckland, New Zealand
m.ghafari@auckland.ac.nz

*Abstract*—WebAssembly is increasingly used as the compilation target for cross-platform applications. In this paper, we investigate whether one can rely on the security measures enforced by existing C compilers when compiling C programs to WebAssembly. We compiled 4,469 C programs with known buffer overflow vulnerabilities to x86 code and to WebAssembly, and observed the outcome of the execution of the generated code to differ for 1,088 programs. Through manual inspection, we identified that the root cause for these is the lack of security measures such as stack canaries in the generated WebAssembly: while x86 code crashes upon a stack-based buffer overflow, the corresponding WebAssembly continues to be executed. We conclude that compiling an existing C program to WebAssembly without additional precautions may hamper its security, and we encourage more research in this direction.

## I. INTRODUCTION

WebAssembly is a recent standard for portable binary code that aims to bring native speed to programs that run in web browsers [7]. Already supported by all major browsers, its use has expanded beyond the web to IoT systems and cross-platform desktop applications [1], [8]. It is now supported as a compilation target by compilers such as Clang and GCC, enabling the compilation of a wide variety of source languages to WebAssembly. The standard has been designed with security in mind, as evidenced among others by the strict separation of application memory from the execution environment's memory. Thanks to this separation, a compromised WebAssembly binary cannot compromise the browser that executes the binary [2], [7].

At first sight, it seems that one can seamlessly enable the cross-platform deployment of a C application by compiling it to WebAssembly. However, the compilers targeting WebAssembly are not yet as mature as for other compilation targets. From a security perspective in particular, it is not yet clear whether the semantics of programs developed in a source language will remain the same when compiled to WebAssembly.

In this paper, we investigate whether we can observe any difference in the outcome of the execution of C programs compiled to native code and to WebAssembly code. We refer to these differences as *divergences in the execution* of the programs. We are particularly interested in determining whether one can rely on the security enforcements seen in existing compilers for native code when compiling to WebAssembly code. To this end, we compiled 4,469 C programs, suffering from known buffer overflow vulnerabilities to x86 executables and WebAssembly executables. We ran these programs and investigated whether there are any divergences in their execution, meaning that the behaviour of the native executable differs from the WebAssembly one. Even if in theory, the presence of undefined behaviour in C program can result in any compiled code, we expect that the compiler protections remain the same across compilation targets. Such divergences can pose a security risk to the deployment of WebAssembly binaries ported from C applications.

We manually investigated the divergent cases and established that they are related to a missing security measure such as stack canaries in the WebAssembly executable. Therefore, a program that is protected when compiled to native code may become insecure when compiled to WebAssembly. This contradicts the design document of the WebAssembly standard [2] which states "*common mitigations such as data execution prevention (DEP) and stack smashing protection (SSP) are not needed by WebAssembly programs*". We discuss the circumstances under which stack smashing may go undetected during the execution of a WebAssembly executable, and we illustrate the impact of compiler optimisations and variations in the code on this behaviour. In summary, this work makes the following contributions:

- We show that C programs that are exposed to stack-based buffer overflows are not protected similarly when compiled to native code versus WebAssembly.
- We discuss a set of examples exposing divergence of behaviour between compilers targeting native x86 code and WebAssembly.
- We share the dataset of 1,088 C programs that exhibited divergent behaviour, together with the corresponding x86 and WebAssembly executables.[1]

This work has three important implications. Firstly, it alerts practitioners that the regular security protections put in place by established compilers may be absent when targeting WebAssembly. Secondly, it encourages researchers to explore whether divergent behaviour exists between a WebAssembly executable and its native counterpart resulting from the compilation of other source languages than C. Thirdly, it encourages researchers to explore possible divergent behaviours that can exist beyond protection against stack-based buffer overflows.

---
[1]https://figshare.com/articles/dataset/QRS_2021_Dataset/16860388

The remainder of this paper is organised as follows. In Section II, we briefly provide the necessary background information on the WebAssembly language and its security model. Section III describes the selection of the dataset that we have used for this research. Section IV details our approach to identify and investigate divergence in the execution of WebAssembly and native programs. The results of our investigation are detailed in Section V, where we illustrate a number of examples exhibiting a divergence and explain the root causes of such divergences in terms of the produced native and WebAssembly binary code. Section VI discusses the impact of compiler optimisations and of differences among examples from our dataset on the divergences. In Section VII, we explain the threats to validity of this work. We discuss related work in Section VIII before concluding the paper in Section IX.

## II. BACKGROUND

We first provide some background information on the WebAssembly language to streamline the understanding of what we discuss in this paper. We focus on the parts of the language that are relevant to our discussion. Readers familiar with WebAssembly can safely skip this section. For a full reference on the language, we refer to its specification [15].

### A. Execution and Memory Model

The execution model of WebAssembly is stack-based: instructions push and pop values on and from the stack. The return value of a function is the top value of the stack after executing the instructions in its body. A WebAssembly program also contains a single *linear memory*, i.e., a consecutive sequence of bytes that can be read from and written to by specific instructions. Using this linear memory properly is left to the program at hand.

### B. Program Structure

WebAssembly programs are composed of, among other components, a number of functions. Functions have a number of parameters, and their body comprises a set of instructions. Some functions can be *exported*, and one such function can be specified as the entry point of the program. Exported functions are made available to the runtime. For example, programs compiled with the WebAssembly System Interface (WASI) expose a `_start` function, corresponding to the `main` function of a C program. This function will be called to start the program.

Functions have access to *local* variables, akin to local registers, in which they can store intermediary results. Function arguments can also be retrieved through these local variables.

An example function is the following:

```
1 (func $main (type 4)
2     (param i32 i32) (result i32)
3     (local i32)
4     local.get 0)
```

This function takes two 32-bit integers as parameters (shown in red), and returns a 32-bit integer (shown in green). When the function is called, the stack is initially empty and the parameters are stored in local variables 0 and 1. A third local variable is accessible as the function declares the need to access an extra local variable on line 3, shown in blue. The function body consists of a single instruction on line 4 (shown in orange). This instruction puts the value of the first local variable (i.e., the first argument of the function) and pushes it on the stack. After the last instruction, the function execution ends and the value remaining on the stack is the return value.

### C. Example Instructions

We have seen the `local.get` instruction to access a local variable and push it on the stack. There exist many other instructions; those of interest are the following:

- `i32.const N` pushes constant N on the top of the stack.
- `i32.add` and `i32.sub` respectively add and subtract the two top values of the stack.
- `local.set N` pops the top value of the stack and sets the $N^{th}$ local variable to this value. `local.tee N` works similarly, but leaves the stack untouched.
- `i32.store` takes two values from the stack and stores the first value at the address pointed to by the second value in the linear memory.

Finally, note that there are variations of these instructions such as `i64.const`, which pushes a 64-bit integer value on the stack, or `i32.store8 offset=N` which stores a byte in the memory with the given offset.

### D. Stack Memory Management

WebAssembly programs also have access to a number of *global* variables, which are accessed and modified similarly to local variables, but through the `global.get` and `global.set` instructions. Most compilers, including Clang and GCC, rely on the first global variable, which we call `g0`, to model the stack memory of the C program they compile using the linear memory of WebAssembly: `g0` acts as the stack pointer. The portion of the linear memory starting at `g0` is therefore used to store stack-allocated data. A common pattern encountered in WebAssembly programs compiled from C that need to allocate space for stack data is therefore the following:

```
1 global.get 0
2 i32.const 64
3 i32.sub
4 global.set 0 ;; g0 becomes g0-64
5 [...] ;; function body
6 global.get 0
7 i32.const 64
8 i32.add
9 global.set 0 ;; restore value of g0
```

This excerpt takes the current stack pointer (`global.get 0`), decreases it by 64, and updates the stack pointer (`global.set 0`). This effectively allocates 64 bytes of data onto the stack. The rest of the function body can therefore store data in that portion of the memory. When the allocated memory is not needed anymore, the stack pointer is restored to its original value (lines 6-9).

*E. Security Model*

The security model of WebAssembly is focused on the fact that a vulnerable WebAssembly program cannot escape the sandbox in which it is executed, thereby avoiding compromising the host running the WebAssembly program. Moreover, multiple features of WebAssembly render programs less vulnerable than their native equivalent. Unlike in x86, the return address of a function in WebAssembly is implicit and can only be accessed by the execution environment, preventing return-oriented programming attacks among others, diminishing the potential of stack smashing attacks. Also, function pointers are supported in WebAssembly through *indirect calls*, where the target function of the call is contained in a statically-defined table: this reduces the number of possible control flow exploits. However, as soon as one sensitive function can be the target of a indirect function call, it is still possible for a vulnerable executable to see its control-flow redirected to call it with untrusted data [11].

Although mechanisms such as stack smashing protections are not needed to protect the host of the execution environment due to the sandboxed nature of WebAssembly, a WebAssembly program may still be prone to vulnerabilities that could allow arbitrary code execution of JavaScript for example [11]. This is not a concern that can be handled by the runtime or sandbox: once the WebAssembly code has been produced by the compiler, no WebAssembly runtime can distinguish between legitimate modifications to the linear memory and attempts at smashing the stack, as the program stack is accessed as any other location of the linear memory.

## III. DATASET

We rely on the Juliet Test Suite 1.3 for C[2] of the *Software Assurance Reference Dataset* [3], released in October 2017 and which has been used to compare static analysis tools that detect security issues in C and Java applications [5]. In particular, we investigate examples that exhibit two common weaknesses: CWE 121 (*stack-based buffer overflow*), and CWE 122 (*heap-based buffer overflow*). This dataset contains thousands of test cases that demonstrate such weaknesses in C programs. Each example contains a vulnerability, but no attempt to exploit it: for example, there is no attempt to reroute the control flow of the program by manipulating function pointers.

*A. Categorisation of Test Cases*

The test cases from this dataset are grouped within categories: all test cases within one category rely on the same mechanism to trigger the weakness, but test cases within one category differ slightly in their control flow. For example, the test case called `dest_char_alloca_cpy_01.c` contains a stack-based buffer overflow, triggered by using a `char` buffer allocated with `alloca` and copied into with

[2]https://samate.nist.gov/SARD/testsuite.php

`strcpy`. This test case has 50 variations, each identified by a different number suffix and demonstrating different usages of the same functions to trigger the stack-based buffer overflow. One variation, for instance, features a more complex control flow stemming from additional conditionals. For our manual inspection, we will group test cases in categories so that only one test case is inspected per category. In total, there are 184 categories across both weaknesses.

*B. Compilation of Test Cases*

Each of the test cases can be configured to exhibit safe or unsafe behaviour. Because we rely on program crashes to identify divergent behaviour, we configured the dataset to only include the unsafe behaviour when compiled.

We compiled each test case to WebAssembly and to x86 with Clang v11.1.0 with the default flags, which include protection against stack smashing attacks through stack canaries[3], the latest version available when we performed our analysis in March 2021, and used a moderate level of optimisations (`-O1`). This level of optimisation was chosen due to the nature of the dataset, where many programs would otherwise have their vulnerable code optimised away, both in their native and WebAssembly version.

We filtered out programs that could not be compiled to WebAssembly, as some programs depend on system-specific functions or on features that are not yet supported by WebAssembly, such as threads or sockets. In total, for CWE 121, we have 2785/5906 programs (47%) that can be compiled to WebAssembly, and for CWE 122, we have 1666/3656 programs (46%) that can be compiled. These programs form our initial dataset.

## IV. DIVERGENCE DETECTION

We describe the process that we followed in order to analyse the programs in our dataset. We also provide a summary of our results.

*A. Identifying Divergence in Execution*

For each program in our dataset, we produced two executables: one compiled to 64-bit x86 native code, and one compiled to WebAssembly. We ran the two executables for each program 100 times with a 5 seconds timeout, and recorded the outcomes including any crashes. If an inconsistency between the two executables was discovered, we reran them manually to confirm the divergence. The programs were executed on a machine running Linux 5.12.9, and the WebAssembly programs were run with Wasmer 1.0.2[4] WebAssembly runtime.

The results of our preliminary experiments are reported in Table I. Even though all compiled test cases exhibit a vulnerability that is triggered during their execution, we see that most executions did not result in a crash, meaning that the vulnerability went unnoticed. In some cases, the WebAssembly and the native executables both crash. This can be either due to its state becoming corrupted so that the execution can no longer be continued, for example the program data has been corrupted and an invalid pointer is accessed; or in the case of native code a protection mechanism can prevent the execution from running to completion. Because there is no divergence in the behaviour of such programs when compiled to WebAssembly or native code, we do not further investigate these cases.

[3]All Clang flags and their default values are listed in the relevant documentation page: https://clang.llvm.org/docs/ClangCommandLineReference.html
[4]https://wasmer.io/

However, we do notice a non-negligible subset of test cases for which only the native executable crashes, respectively 31% and 14% for CWE 121 and CWE 122. This indicates that the behaviour of the WebAssembly executable does not match the behaviour of the native one. We make this dataset of C sources, WebAssembly binaries, and x86 binaries public[5].

TABLE I
DIVERGENCES BETWEEN WEBASSEMBLY AND NATIVE EXECUTABLES.

| CWE | No crash | Both crash | Wasm crash | Native crash |
|---|---|---|---|---|
| CWE 121 | 1824 | 107 | 0 | 854 |
| CWE 122 | 1342 | 108 | 0 | 234 |

### B. Sanity Checks

Before inspecting the executables that have diverging behaviour manually, we performed two automated sanity checks:

- *Crash report:* We looked at the actual crash report for the executed binary. In all divergent cases, the execution triggered a protection mechanism against stack smashing, recognisable by the following output:

  ```
  *** stack smashing detected ***:
     terminated
  ```

- *Compiler dependence:* We recompiled each program using GCC v11.1.0 instead of Clang to see whether the native x86 code results in the same crash. In all divergent cases, the native code produced by GCC resulted in a `SIGSEGV` error. We manually inspected the produced binaries to confirm that GCC also introduced stack canaries.

### C. Manual Inspection

The 1,088 test cases that exposed a divergent behaviour belong to 70 different categories from the test suite. Within one category, each program differs only slightly in their control flow, but rely on the same standard library functions to trigger the vulnerability. Hence, within each category, we manually investigated one of the program that expose a divergent behaviour. We cover the differences that arise within a category in Section VI-B. Our manual investigation entailed the following.

- *Root cause of the crash*: by inspecting the C code, we determine whether the program is supposed to crash, and track down its root cause. We noticed that even though the CWE 122 examples are supposed to demonstrate heap-based buffer overflows, all divergent cases actually crash due to a stack-based buffer overflow.
- *Inspecting the decompiled x86 and WebAssembly binaries*: by manually inspecting these binaries, we further pinpoint the root cause for the divergence

## V. OBSERVATIONS

All the 1,088 divergences that we encountered are related to stack-based buffer overflows. The non-crashing behaviour of the WebAssembly version of the executables is due to the absence of a protection mechanism like stack canaries in the generated code. We manually inspected one program that expose divergent behaviour for each of the 70 categories. We detail the analysis of one program exposing divergent behaviour of each dataset (CWE 121 and CWE 122).

### A. Stack-Based Buffer Overflow (CWE121)

Consider the following excerpt from the example `CWE805_char_declare_loop_05` of the CWE 121 dataset:

```
1  char * data;
2  char dataBadBuffer[50];
3  data = dataBadBuffer;
4  data[0] = '\0';
5  char source[100];
6  memset(source, 'C', 100-1);
7  source[100-1] = '\0';
8  for (i = 0; i < 100; i++) {
9      data[i] = source[i];
10 }
11 data[100-1] = '\0';
12 printLine(data);
```

This code allocates a destination buffer of 50 elements in the stack memory on line 2, and a source buffer of 100 elements on line 5, before copying the entire contents of the source buffer to the destination buffer through the loop on line 8. However, because the destination buffer is too small, elements will be copied outside of the destination boundary.

We perform a manual inspection of the native x86 executable with `radare2`[6]. It operates as one would expect from the C source code, but it contains the following addition at the end of the compiled function:

```
mov rax, qword fs:[0x28]
mov rcx, qword [var_8h]
cmp rax, rcx
jne 0x12b7
add rsp, 0xc0
pop rbp
ret
call sym.imp.__stack_chk_fail
```

This is the code that checks the stack canary generated by the compiler and stops the program's execution when a stack overflow is detected. The canary is the value of the stack pointer before the execution of the function, and it is compared to the stack pointer after the execution of the function. If an inconsistency is detected, the execution of the program is aborted by calling __stack_chk_fail.

---

[5]URL redacted for blind review

[6]https://rada.re/n/

The C excerpt compiles to the following WebAssembly:

```
1  (func $main (type 4)
2    (param i32 i32) (result i32)
3    (local i32)
4    global.get 0
5    i32.const 64
6    i32.sub
7    local.tee 2 ;; l2 = g0-64
8    global.set 0 ;; g0 = g0-64
9    i32.const 0
10   ...
11   local.get 2 ;; [g0]
12   i32.const 67 ;; ['C', g0]
13   i32.const 99 ;; [99, 'C', g0']
14   call $memset
15   local.tee 2
16   i32.const 0
17   i32.store8 offset=99
18   ...)
```

We notice that the source buffer has been inlined; the generated code operates directly on the destination buffer. Line 8 moves the stack pointer (g0) up by 64 bytes to allocate stack space for the destination buffer. The memset call on line 14 fills it with 99 C characters. Afterwards, line 17 writes the final null character to position 99 of the destination buffer. Note that because the destination buffer has only been allocated 64 bytes, these operations will result in a stack overflow. However, we do not observe the presence of a stack protection mechanism in the WebAssembly code, and executing the WebAssembly program will not result in a crash, letting the stack overflow occur silently.

### B. Heap-Based Buffer Overflow (CWE122)

Consider the following excerpt from the c_src_wchar_t_cat_03 example of the CWE 122 dataset:

```
1  wchar_t * data;
2  data = (wchar_t *)malloc(
3      100*sizeof(wchar_t));
4  if (data == NULL) {exit(-1);}
5  wmemset(data, L'A', 100-1);
6  data[100-1] = L'\0';
7  wchar_t dest[50] = L"";
8  wcscat(dest, data);
9  printWLine(data);
10 free(data);
```

Even though this example is part of the heap-based buffer overflow dataset, it contains a stack-based buffer overflow that is similar to the previous example. A source buffer of 100 wide characters is allocated on line 3 and filled on line 5. Then, a destination buffer of 50 wide characters is allocated on line 7, and the source buffer is copied into it on line 8. However, because the source buffer is larger than the destination buffer, the extraneous data will be copied outside of the destination buffer and will overflow the stack. The WebAssembly and native code for this example are similar to the previous example, and we therefore do not detail them further.

### C. Discussion

From our observation, we draw the conclusion that WebAssembly binaries are indeed vulnerable to stack smashing attacks. This is mitigated by the fact that the design of WebAssembly avoids control flow hijacking attacks through e.g., return oriented programming and that WebAssembly programs are executed in a sandboxed environment, but this contradicts the design documents stating that stack smashing protections are not needed for WebAssembly. Moreover, stack smashing protection need to be ensured by the compiler rather than the WebAssembly runtime, that has no knowledge of how the program stack is managed.

Another important aspect is that the vulnerable code discussed here relies on undefined behaviour. Even though this means that compilers are allowed to transform programs with buffer overflows as they see fit, in accordance with the principle of least astonishment, it is reasonable to expect that the same protection mechanisms are used across all compilation targets.

The divergences exposed here therefore mean that the WebAssembly binaries are exposed to stack-smashing attacks in the sense that, even though the program remains sandboxed in its environment, the control or data flow of the program may be exploited by an attacker.

## VI. MISCELLANEOUS

We now touch upon two other concerns regarding our setup for the evaluation: the impact of the optimisation level, and the differences that can occur within variants among one category.

### A. Impact of Optimisations

We now turn our attention to the impact of optimisations on the ability to prevent vulnerable code to make it into the binary[7]. To illustrate this, we inspect one specific example that exhibits different behaviour depending on the optimisation levels, which is in essence similar to our first example. This example contains a stack-based buffer overflow.

```
int * data = (int *)alloca(10);
int source[17] = {0};
for (size_t i = 0; i < 17; i++)
{
    data[i] = source[i];
}
printIntLine(data[10]);
```

---
[7]Compiling without optimisation is not desirable if one wants to model real-world scenarios.

*a) Compiled with -O0:* Compiled without optimisations (using the -O0 flag), the native binary does *not* crash, indicating that the stack canary did not detect the stack overflow.

*b) Compiled with -O1:* Compiled with the -O1 flag, the native binary does crash upon the detection of the stack overflow. The WebAssembly binary suffers from the same stack overflow. On line 5, 16 bytes are allocated on the stack; yet on line 10 a value is copied 32 bytes beyond the stack pointer. Yet, this binary continues executing without crashing.

```
1  global.get 0
2  i32.const 16
3  i32.sub
4  local.tee 0
5  global.set 0
6  local.get 0
7  i32.const 32
8  i32.add
9  i64.const 0
10 i64.store
11 ...
```

*c) Compiled with -O2 and -O3:* Compiled with -O2 and -O3, the native binary does not crash anymore. Inspecting the native code, we see that the argument to the printIntLine call has been inlined, which eliminates the overflow:

```
mov qword [arg_20h], 0
mov edi, dword [rsp]
call sym.printIntLine
```

In the generated WebAssembly, in contrast, the overflow has not been eliminated by the optimisations even though the argument to printIntLine has also been inlined:

```
global.get 0
i32.const 16
i32.sub
local.tee 0
global.set 0
local.get 0
i32.const 32
i32.add
i64.const 0
i64.store
...
i32.const 0
call 14 ;; printIntLine
```

This suggests that compiling with a lower level of optimization may reveal more differences, although this would not reflect a real-world scenario. For this reason, we decided to use -O2 as the level of optimization.

## B. Variants Among a Category

Finally, we investigated all the variants inside the CWE131_memcpy category of the CWE121 dataset for which the native binary crashes. We picked this category as the original program (variant 01) remains small. This sheds some light on why among one category there can be variants that crash and variants that do not crash.

*a) Variant 01:* The first –and simplest– variant is the following, and its native version does not crash. The root cause for a possible overflow is that the data buffer is allocated 10 bytes, while it should be 40 bytes long to hold 10 integers.

```
int * data = (int *)alloca(10);
int source[10] = {0};
memcpy(data, source, 10*sizeof(int));
printIntLine(data[0]);
```

Looking at both the WebAssembly and the native executables, we see that the argument to printIntLine has been inlined, and that all buffers have been removed, thereby resulting in code equivalent to printIntLine(0).

*b) Variants 02 to 08:* These variants all are variations where the allocation site is enclosed in a conditional that is always true. For example, variant 02 starts as follows:

```
int * data = NULL;
if (1) {
    data = (int *)alloca(10);
}
```

We do not observe divergent behaviours on any of these variants: they all are compiled like variant 01.

*c) Variant 09:* This variant similarly encloses the allocation in a conditional that is always true. However, this time the value of the conditional cannot be known by considering the example's source file in isolation, as it depends on the value of GLOBAL_CONST_TRUE which is defined in another file included during the compilation.

```
int * data = NULL;
if(GLOBAL_CONST_TRUE) {
    data = (int *)alloca(10);
}
```

Looking at the generated WebAssembly and native code, we recognise the GLOBAL_CONST_TRUE global being loaded which is followed by the allocation of the memory and the overflow. We do notice that the argument given to printIntLine is inlined as 0.

*d) Variants 10 to 14:* All these variants differ in the conditional used, which always evaluates to true. In some cases, the conditional depends on a variable, constant, or a function defined in a different file. We list below the relevant lines for each of these variants:

```
if(globalTrue) // Variant 10
if(globalReturnsTrue()) // Variant 11
// Variant 12
if(globalReturnsTrueOrFalse())
if(GLOBAL_CONST_FIVE==5) // Variant 13
if(globalFive==5) // Variant 14
```

All these variants compile similarly to variant 09, and behave the same: the native binary crashes while the WebAssembly binary does not.

*e) Variant 45:* This variant is the only remaining one that exhibits divergent behaviour. Even though the semantics of the C program is similar to the other variants, the vulnerability occurs across two function calls, resulting in different compiled code. One function performs the stack allocation, storing the destination buffer in a shared static variable, and calls another function that performs the offending copy in the destination buffer. Because this vulnerability is spread across two functions using a shared static variable, the optimisations that are applied in the other variants are not applied here, and the vulnerability remains in both the native and the WebAssembly executables. The native executables crashes with a stack smashing detection, while the WebAssembly executables executes to completion.

*f) Other variants:* All other variants did not result in divergent behaviour. This is due to the fact that the compiler was able to optimise the program in a way that renders it equivalent to the 01 variant. For example, variant 17 encloses the allocation of the destination buffer in a for loop that performs a single iteration:

```
for(i = 0; i < 1; i++) {
    data = (int *)alloca(10);
}
```

## VII. THREATS TO VALIDITY

We identify three main threats to validity. First, we rely on an existing test suite of programs with known vulnerabilities. The use of this test suite and our selection programs within this test suite is a threat to external validity. The latest release of the Juliet Test Suite dates from 2017, and there could be other vulnerabilities or patterns to express them that are not present within the test suite. For instance, one aspect which we have not encountered in the programs we inspected is that there is no use of function pointers. We selected all programs that exhibit two common vulnerabilities (CWE 121 and CWE 122). Whether divergent behaviour can also arise for programs with other vulnerabilities is left for future work. Moreover, we only focused on programs that are unsafe: the test cases can also be configured to exhibit safe behaviour only. Divergences in the execution of such safe programs could arise, but their detection requires a more involved analysis. In this work, we rely on program crashes to identify divergences. This means that divergences might have been missed, but that the observed divergences are actual divergences nonetheless.

A threat to internal validity is the fact that we focus on compiler inconsistencies which are exhibited through divergent behaviour. Identifying divergent behaviour for programs that result in the same outcome is something that requires a thorough analysis of programs that we leave for future work.

Finally, another threat to internal validity regards the setup of our evaluation. We performed our evaluation using Clang to compile programs in WebAssembly: while it is also possible to compile them using GCC through the emscripten toolchain[8]. We leave a comparison to the results of GCC for future work. Moreover, all executables were run on a Linux platform with the wasmer runtime, results may differ on another platform or with another runtime, although the root cause of the divergences are linked to the compiler. We performed our evaluation using -O1 as the optimisation level. We have shown that the optimisation level can have an impact on whether the divergence is present in the binary in Section VI-A level of optimisations and compiler flags we used for our evaluation. Repeating the evaluation with various compiler configurations could highlight more root causes of divergence.

[8]https://emscripten.org/

## VIII. RELATED WORK

The design of WebAssembly has been developed with security in mind [2]. Hilbig et al. [9] have studied the current usage of WebAssembly on the web and beyond. The results of this study notably demonstrate that around two thirds of WebAssembly binaries are developed in memory-unsafe languages, and are therefore prone to vulnerabilities that affect the source language.

Despite protective measures such as the isolation of application memory from the runtime environment, WebAssembly binaries may still suffer from a number of weaknesses as identified by Lehmann et al. [10], rendering them easily exploitable —more easily than native binaries. For example, the presence of critical functions exported from the environment such as `eval`, `exec`, or `fwrite` enables arbitrary code execution.

There have been numerous works related to improving the security of WebAssembly these past years. To mitigate the attack vectors on vulnerabilities that a WebAssembly program can have, Arteaga et al. [4] propose an approach to achieve code diversification for WebAssembly: given an input program, multiple variants of this program can be generated. Narayan et al. [14] propose Swivel, a new compiler framework for hardening WebAssembly binaries against Spectre attacks, which can compromise the isolation guarantee of WebAssembly. Stiévenart and De Roover propose a static analysis framework [16] for WebAssembly, used to build an information flow analysis [17] to detect higher-level security concerns such as leaks of sensitive information. Namjoshi et al. [13] introduce a self-certifying compiler for WebAssembly, so that the optimisations performed during compilation are generated with proofs of their correctness. In the examples we have discussed, such proofs would not help: the semantics of the source program is preserved correctly, but the compiler does not introduce the same protection mechanisms as it does for its x86 backend. Disselkoen et al. [6] propose an extension to WebAssembly that allow developers to encode C and C++ memory semantics in WebAssembly, by reifying segments and handles in the language.

Our method for identifying divergence is inspired by related work on compiler fuzzing. Csmith [18] performs fuzzing of compilers by generating C programs, compiling them with several compilers, and observing their result: any difference in output is likely a bug. Similarly, the approach of Midtgaard [12] generates OCaml, that it compiles with two different backends of the compiler: one for native code and one for OCaml bytecode. Any difference in output is flagged as a potential bug. In our case, we also target two backends of the same compiler, but we do rely on a predefined set of input programs by using the Juliet Test Suite for C.

## IX. Conclusion

In order to increase understanding of what can happen when ones ports a C application to WebAssembly, we investigated whether divergences can arise between the execution of an application compiled to WebAssembly and the execution of the same application compiled to another target such as x86. and if they can arise, whether it may have an impact on the security of the application. We have observed that, on a dataset of 4,469 vulnerable C programs containing buffer overflows, 24% of the programs differ in outcome when their WebAssembly and their x86 binary is executed. We manually inspected such divergences and observed that they are due to a lack of security measures such as stack canaries in the generated WebAssembly. Therefore, applications that were protected from certain vulnerabilities when compiled to native code can become insecure when compiled to WebAssembly. We believe that this is an important observation for practitioners and that follow-up research is needed to identify whether more of such divergences can occur. Moreover, research is needed to correct these divergences in order to enable the secure deployment of WebAssembly binaries compiled from C programs.

## Acknowledgement

This work has been partially funded by the Cybersecurity Initiative Flanders.